\documentclass[runningheads]{llncs}
\usepackage[T1]{fontenc}
\usepackage{gensymb}
\usepackage{cite}
\usepackage{graphicx}
\usepackage{tabularx}
\usepackage{booktabs}

\begin{document}
\title{RESISTO Project: Automatic detection of operation temperature anomalies for power electric transformers using thermal imaging}
\titlerunning{Resisto project}
%
\author{David López-García \inst{1} \and Fermín Segovia \inst{1} \and Jacob Rodríguez-Rivero \inst{3} \and Javier Ramírez \inst{1} \and David Pérez \inst{2} \and Raúl Serrano \inst{2} \and Juan Manuel Górriz \inst{1}}
\authorrunning{David López-García et al.}
%
\institute{Department of Signal Theory, Networking and Communications, University of Granada, Spain \\
\and ATIS Soluciones \& Seguridad, Granada, Spain \\
\and e-Distribución Redes Digitales, Spain} 
\maketitle              
\begin{abstract}
The RESISTO project represents a pioneering initiative in Europe aimed at enhancing the resilience of the power grid through the integration of advanced technologies. This includes artificial intelligence and thermal surveillance systems to mitigate the impact of extreme meteorological phenomena. RESISTO endeavors to predict, prevent, detect, and recover from weather-related incidents, ultimately enhancing the quality of service provided and ensuring grid stability and efficiency in the face of evolving climate challenges. In this study, we introduce one of the fundamental pillars of the project: a monitoring system for the operating temperature of different regions within power transformers, aiming to detect and alert early on potential thermal anomalies. To achieve this, a distributed system of thermal cameras for real-time temperature monitoring has been deployed in The Doñana National Park, alongside servers responsible for the storing, analyzing, and alerting of any potential thermal anomalies. An adaptive prediction model was developed for temperature forecasting, which learns online from the newly available data. In order to test the long-term performance of the proposed solution, we generated a synthetic temperature database for the whole of the year 2022. Overall, the proposed system exhibits promising capabilities in predicting and detecting thermal anomalies in power electric transformers, showcasing potential applications in enhancing grid reliability and preventing equipment failures.

\end{abstract}
\section{Introduction}
\subsection{Introduction to the RESISTO project}
The sixth report \cite{IntergovernmentalPanelonClimateChangeIPCC2023} prepared by the Intergovernmental Panel on Climate Change of the United Nations (IPCC) highlights that meteorological phenomena are becoming increasingly extreme and frequent due to global warming. According to the report, these changes in weather patterns are posing significant challenges to communities worldwide. This trend has been the catalyst for the development of the RESISTO project \footnote{Official website: https://www.endesa.com/en/press/press-room/news/energy-sector/endesa-presents-pioneering-project-europe-reduce-negative-effects-weather-power-grid}, which aims to ensure that electrical grids are prepared to mitigate and reduce the impact of these extreme meteorological phenomena.

Under the leadership of Endesa's Grids subsidiary, e-distribución, the RESISTO project stands as a groundbreaking technological innovation and research transfer initiative in Europe. Its primary objective is to enhance the power grid of the Doñana National Park, located in Andalusia, in the south of Spain, through the innovative integration of artificial intelligence and a comprehensive array of cutting-edge technologies. These include thermal surveillance cameras, weather stations, fire sensors, and an autonomous drone fleet. By leveraging these advancements, the RESISTO project aims to mitigate the adverse effects of weather-related phenomena, such as extreme heat, wind, water, and other potential risks on the power grid, which elevates the quality of service provided to the region. To achieve this goal, RESISTO has proposed a pioneering solution focusing on four crucial areas:

\begin{enumerate}
\item Planning: Utilizing AI and Big Data for prediction and prevention, identifying high-risk areas prone to incidents during specific weather conditions.
\item Detection: Monitoring the grid using sensors, thermal cameras, and surveillance equipment to control vegetation growth.
\item Recovery: Employing AI and a fleet of drones for real-time incident location and providing crucial support for operational and maintenance tasks, especially in remote or hard-to-access areas.
\item Adaptation through continuous learning of the machine-learning algorithm implemented, which will be fed with the acquired data to enhance the grid's resilience and efficiency over time.
\end{enumerate}

As the largest electricity company in Spain and one of the largest in the world (serving millions of customers in Spain and several other countries), Endesa and all the public and private institutions involved in the project, bring unparalleled expertise to the endeavor.

\subsection{Mitigating transformer risks in electricity networks}

The oversight of electricity networks demands meticulous consideration towards potential overloads \cite{Vitolina2015,Mullerova2015}, given their prominence as a significant source of issues impacting power transformers, responsible for approximately 12\% of total failures \cite{Yazdani-Asrami2015}. Specially, transformer malfunctions pose a considerable hazard, capable of igniting fires that, given the urban positioning of transformers, can lead to severe economic and personal ramifications \cite{dolata2016online}. In recent years, different reports examining the likelihood of failure in power transformer components have been published by various authors and international organizations \cite{Barkas2022,AJ2018,Haghjoo2016,Setayeshmehr2004}. Among these organizations, the International Council on Large Electric Systems (CIGRE) has categorized failure types into six distinct categories based on the component responsible for the failure \cite{Abbasi2022}: winding (38\%), tap changer (31\%), bushing (17\%), auxiliary (11\%) core (1\%) and tank (1\%). In order to mitigate these failures, beyond routine maintenance of installations, a series of measurement should be conducted regularly. These, among others \cite{Bakar2014} encompass monitoring operational temperatures \cite{Kunicki2020}. Consequently, the measurement of both transformer and ambient temperature constitutes a fundamental component of standard protocols for power transformer monitoring \cite{Peimankar2017,Velasquez-Contreras2011}. These parameters are pivotal variables that necessitate accurate and uninterrupted measurement and can further serve as predictors for other variables, such as reactive power or current intensity \cite{ramirez2020power}. \\

Early failure prevention techniques in the electrical grid, as well as a significant reduction in associated risks, are closely linked to artificial intelligence techniques that allow us to identify and notify of these failures. Artificial intelligence, and more specifically, big data and machine learning, have made a strong impact in various sectors of society in recent years \cite{Gorriz2020, Gorriz2023} This includes several and diverse areas such as autonomous transportation \cite{Ma2020}, neuroscience\cite{Lopez-Garcia2022, Lopez-Garcia2020, PENALVER2023119960}, healthcare\cite{Lopez-Garcia2018,Chen2019} or automatic text translation \cite{Mohamed2024}. The management of the electrical grid has also been heavily influenced by the emergence of artificial intelligence. For example, these techniques provide the electrical grid with the capability to make intelligent decisions in response to sudden changes in customer energy demand, abrupt increases or decreases in renewable energy production, or extreme weather phenomena. With the increase in data volume, it is also possible to employ machine learning for the detection and prevention of anomalous behaviour \cite{Azad2019, Hossain2019}. \\

In this article, we describe and discuss the development of one of the core modules of the RESISTO project: the monitoring of the thermal performance of power electric transformers. The proposed solution integrates a distributed network of thermal cameras strategically positioned to capture real-time thermal data across various installations. To achieve this, we employ adaptive machine learning algorithms commonly used in time series forecasting, enabling us to predict the expected thermal behavior of the transformer. Consequently, this facilitates the real-time detection and reporting of potential thermal anomalies.

\section{Materials and methods}

\subsection{Thermographic data acquisition}
Temperature measurement of power transformers typically involves the use of sensors that are integrated into the transformer equipment, serving, to some extent, as components of the transformer system. While this method may be the most uncomplicated and direct, it carries the drawback of being susceptible to potential transformer issues, resulting in a loss of its predictive capability for identifying transformer malfunctions \cite{DeMelo2021} . Contrary, thermal imaging, which is a well-established technology that enables the detection of the heat emitted by objects with temperatures above absolute zero \cite{Mariprasath2018}, presents several benefits. The most crucial is the system's complete independence from the power transformer, operating without direct physical contact, thereby avoiding susceptibility to potential transformer failures (such as high temperature peaks or excessively frequent temperature fluctuations that may harm the sensors). Moreover, it proves to be a more scalable system, capable of monitoring not only the overall temperature of the transformer but also distinct components independently or adjacent equipment. This technology has proven effective in various applications, such as monitoring civil structures \cite{Sirca2018, Lu2021}, inspecting machinery \cite{bagavathiappan2008condition}, and addressing temperature issues in the nuclear industry \cite{Itami2004}. Additionally, it has recently been suggested for the correction and monitoring of temperature-related issues in electrical substations and power transformers \cite{Zarco-Perinan2021, Segovia2023, martinez2019prediction}. 

\begin{figure}
\includegraphics[width=\textwidth]{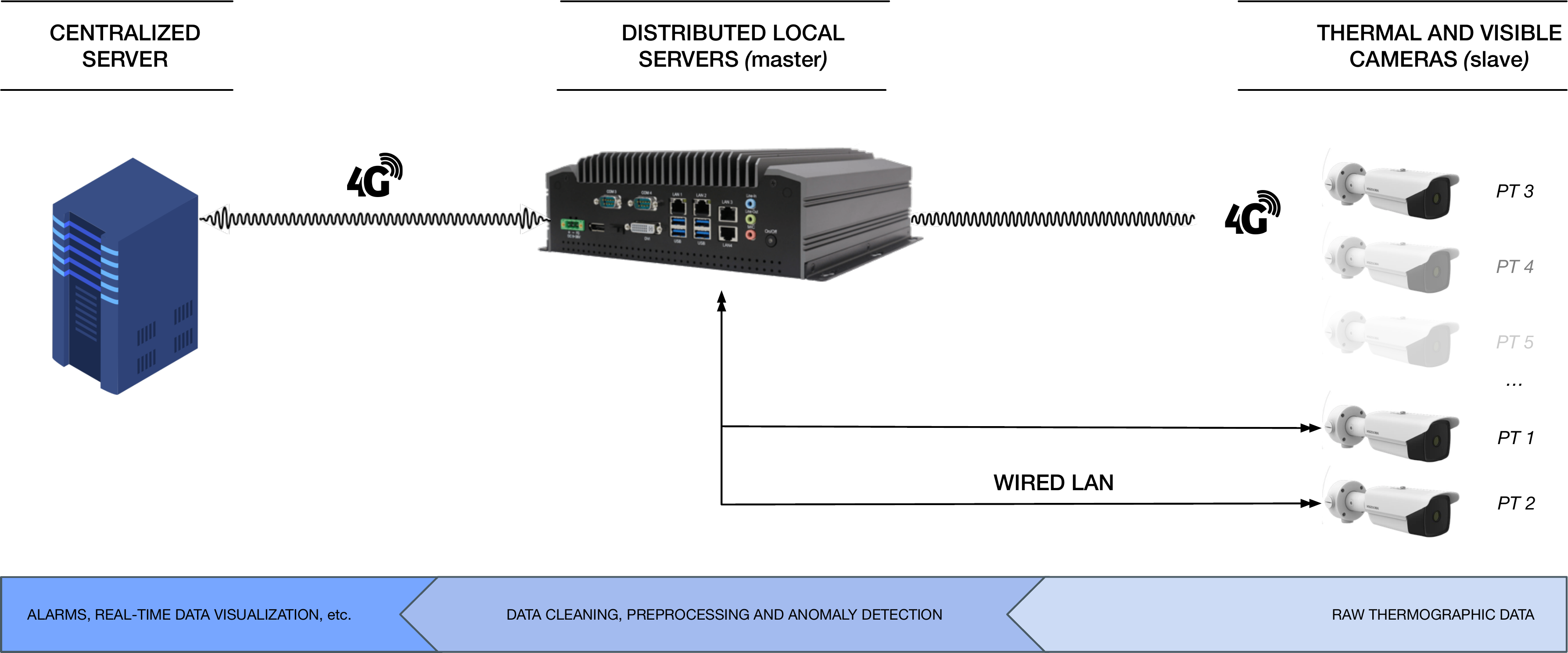}
\caption{Simplified diagram of the thermographic data acquisition system.} \label{architecture}
\end{figure}

\subsubsection{System architecture} The thermography data acquisition and processing system comprises 20 thermal cameras and 9 industrial PCs controlling them (Figure \ref{architecture}) The technical specifications of the diverse camera models and industrial PCs are detailed in the subsequent sections. Industrial PCs establish connections with the cameras in two distinct manners. In cases where the camera coexists in the same power control center as the industrial PC, both are physically linked through an Ethernet cable. Alternatively, when the camera is situated at a different location than the controlling industrial PC, they establish communication through a 4G communication. The radio-link connections are secured using a VPN to adhere to and ensure project security requirements. While contemplating connectivity options, the notion of utilizing radio-links was considered. However, this idea was dismissed due to inadequate coverage in several deployment sites.

The capture of thermal images is initiated upon request from the industrial PC. The acquisition frequency is contingent upon the type of connection established between the industrial PC and the camera (one image per minute for physical links and one image every 5 minutes for radio links). To achieve this, the industrial PC connects to the camera using TCP network protocol (Transmission Control Protocol) and requests the image. Once received, the industrial PC is responsible for its storage and processing, as described further in subsequent sections. To maintain meticulous control over the acquisition time and ensure a simple and scalable data management approach, the images are stored in individual files with filenames based on the acquisition timestamp, following the format \texttt{yyyy-MM-dd HH:MM:SS}.

Ultimately, in order to perform different tasks such as monitoring, control, or data extraction, the industrial PCs are equipped with 4G connectivity, enabling external access using the SSH protocol. Similar to the aforementioned radio-links, these communications are also safeguarded by VPN for enhanced security.

\subsubsection{Thermographic cameras} At the moment, three different models of thermographic cameras were employed for monitoring the temperature of twenty power transformers, two power lines and their surroundings. The specifications of the different models are described below:

\paragraph{Hikvision Thermographic Network Bullet Camera: } This particular model (DS-2TD2137T-4/P) operates at a frame rate of 50Hz and provides thermal images with a resolution of 384x288 pixels. It has a field of view (FOV) of 90\degree x 65.3\degree and a focal length of 4.4mm, effectively covering the transformer's body and its surroundings. The Noise Equivalent Temperature Difference (NETD) is $\leq$35mK and the temperature interval of detection ranges from -20\degree to 550\degree with an accuracy of $\pm$2\degree. It supports several network protocols including IPv4/IPv6, TCP, UDP, etc and different security protocols such as user authentication (ID and PW), MAC address binding, HTTPS encryption, among others. 

\paragraph{Hikvision Bi-spectrum Thermography Network Bullet Camera:} This model (DS-2TD2628T-3/QA) presents two acquisition modules, one thermal and one optical. The thermal module operates at a frame rate of 50Hz and provides thermal images with a resolution of 256x192 pixels. It has a FOV of 50\degree x 37.3\degree, a focal length of 3.6mm, the NETD is $\leq$ 40 mK and the temperature interval of detection ranges from -20\degree to 550\degree with an accuracy of $\pm$2\degree. The optical module provides images with a resolution of 2688x1520 pixels. It has a FOV of 84\degree x 43.1\degree and a focal length of 4.3mm and supports the same network and security protocolos as the previous models. This camera model was used to monitoring both power transformers, power lines and their surroundings.

\paragraph{Sunell Temperature Alarm Bullet Network Camera:} This model (SN-TPC6401KT-F II) operates at a frame rate of 50Hz providing thermal images with a resolution of 640x512 pixels and different focal lengths (15, 25, 35 and 50 mm). The FOV depends on the selected focal length. The (NETD) is $\leq$40mK and the temperature interval of detection ranges from -20\degree to 150\degree with an accuracy of $\pm$2\degree. It also supports several network and security protocols.

\paragraph{Hikvision Network IR Speed Dome:} Finally, we installed a PTZ (Pan, Tilt and Zoom) camera model (DS-2DF8425IX-AELW(T5)) in a specific power line not for monitoring temperatures but to meet other requirements of the project such as the control of the avifauna. These type of cameras are usually employed for monitoring wide areas in high-definition, such as rivers, forests, roads, etc. 

\begin{figure}
\includegraphics[width=\textwidth]{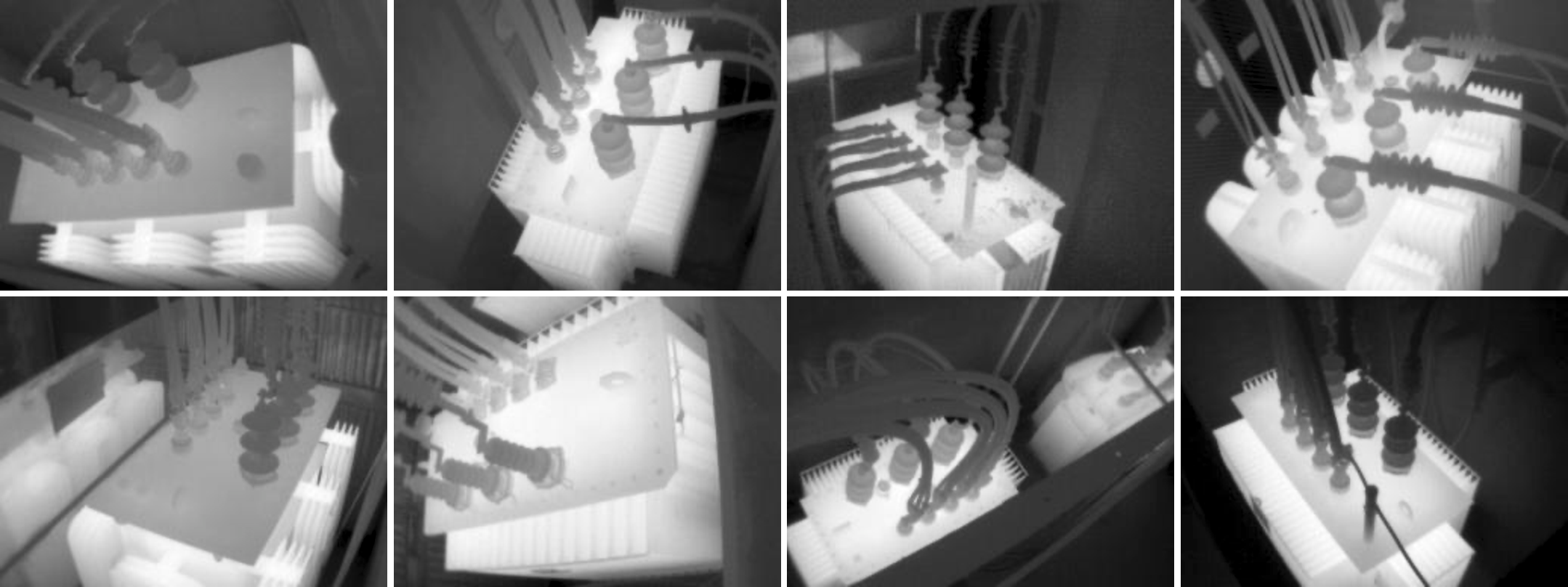}
\caption{Raw thermographic images obtained by DS-2TD2137T-4/P cameras in eight randomly selected power transformation centers.} \label{thermal_cameras}
\end{figure}

\subsubsection{Industrial PCs} Four industrial PCs (TBOX-2825) are collecting, storing and analyzing the thermographic data recorded by the cameras. The equipment has been chosen to ensure the proper functioning of the system under adverse weather conditions, given that the locations where they will be deployed may experience extreme temperatures in both summer and winter. This model can operate within a temperature range of -20\degree to 70\degree, with humidity levels between 10\% and 95\%, and ensures storage integrity within a range of -40\degree to 80\degree. It features an Intel Whiskey Lake 8th Gen Core i5-8265U CPU (Quad-core 1.6GHz, burst to 3.9GHz, TDP 15W), single DDR4 memory (16GB), and 2.5” SATA storage (4TB).

\subsection{Thermal anomalies detection system} 
\begin{figure}
\includegraphics[width=\textwidth]{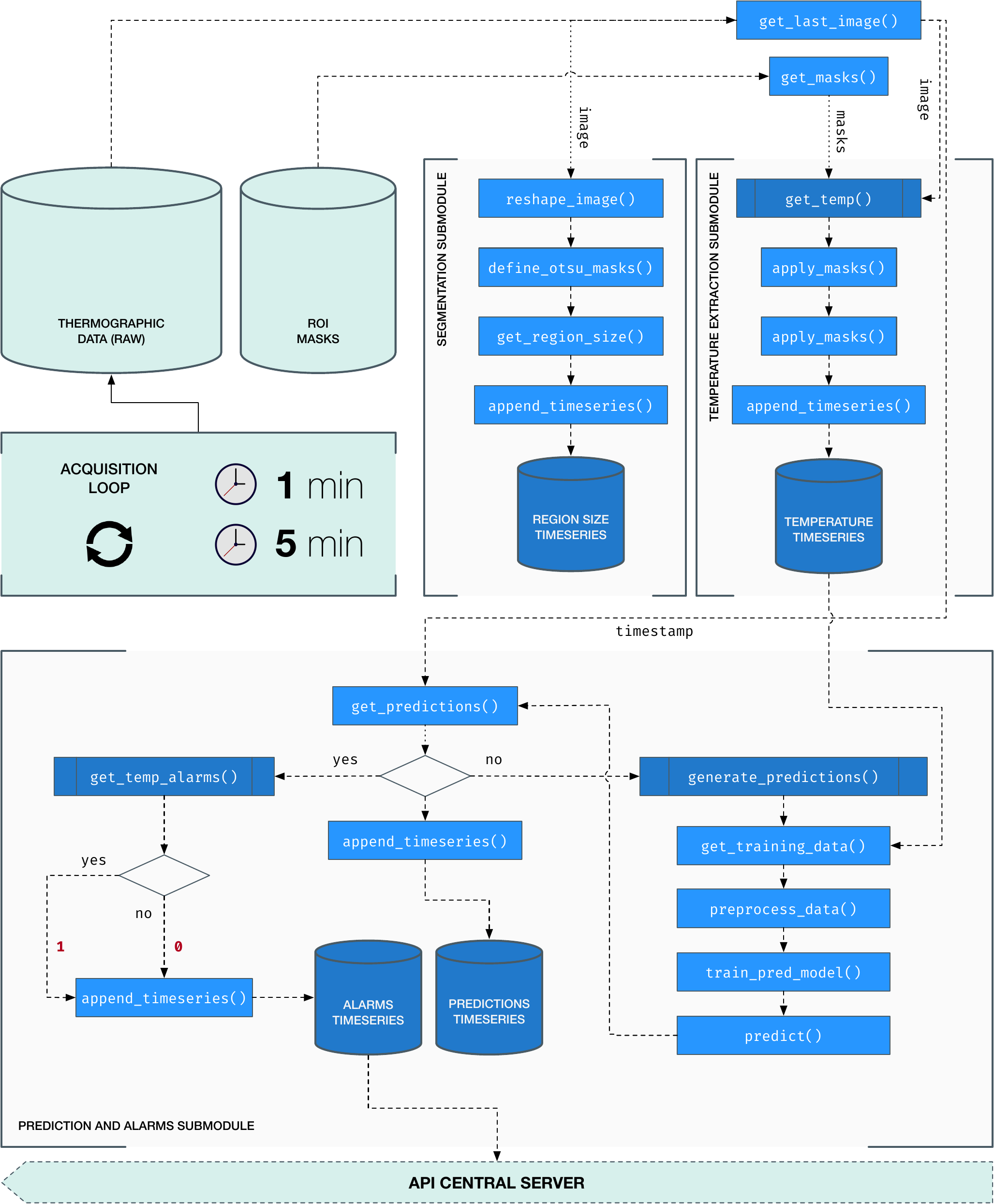}
\caption{Flowchart of the proposed thermal anomaly detection system.} \label{diagram}
\end{figure}

This section outlines the general operation of the proposed thermal anomaly detection solution. This program runs continuously on each of the distributed servers and is independent of the thermal image acquisition system defined earlier. When a thermal image is captured by the camera and received on the server, the latter processes this image to verify if the resgistered temperature of each monitored region of the transformer falls within optimal margins estimated by an adaptive prediction model. If not, a thermal anomaly alarm is generated. The flowchart of the proposed solution is depicted in Figure \ref{diagram}. Below, we provide a more detailed description of each module within the system.

\subsubsection{Thermographic data processing} The thermal image processing is managed by the segmentation and temperature extraction submodules. Image segmentation is carried out in two different ways. Firstly, through an automated segmentation process, and secondly, by manually-defined masks established a priori. Both the manual and the automatic segmentation processes are described below. 

\begin{table}[]
\caption{Nine selected regions of interest and their labels}
\begin{tabular*}{\textwidth}{@{\extracolsep{\fill}}ll@{}}
\toprule
\multicolumn{1}{l}{\textbf{Region of interest (ROI)}} & \multicolumn{1}{r}{\textbf{ROI label}} \\ \midrule
Primary terminal (1) and high voltage bushing  & \texttt{in\_1} \\
Primary terminal (2) and high voltage bushing  & \texttt{in\_2} \\
Primary terminal (3) and high voltage bushing  & \texttt{in\_3} \\
Secondary terminal (1) and low voltage bushing  & \texttt{out\_1} \\
Secondary terminal (2) and low voltage bushing  & \texttt{out\_2} \\
Secondary terminal (3) and low voltage bushing  & \texttt{out\_3} \\
Secondary terminal (4) and low voltage bushing  & \texttt{out\_4} \\
Transformer body  & \texttt{body} \\
Background (building walls, fences, etc.)  & \texttt{background} \\
	\bottomrule
\end{tabular*}
\label{rois}
\end{table}

\begin{figure}
\includegraphics[width=\textwidth]{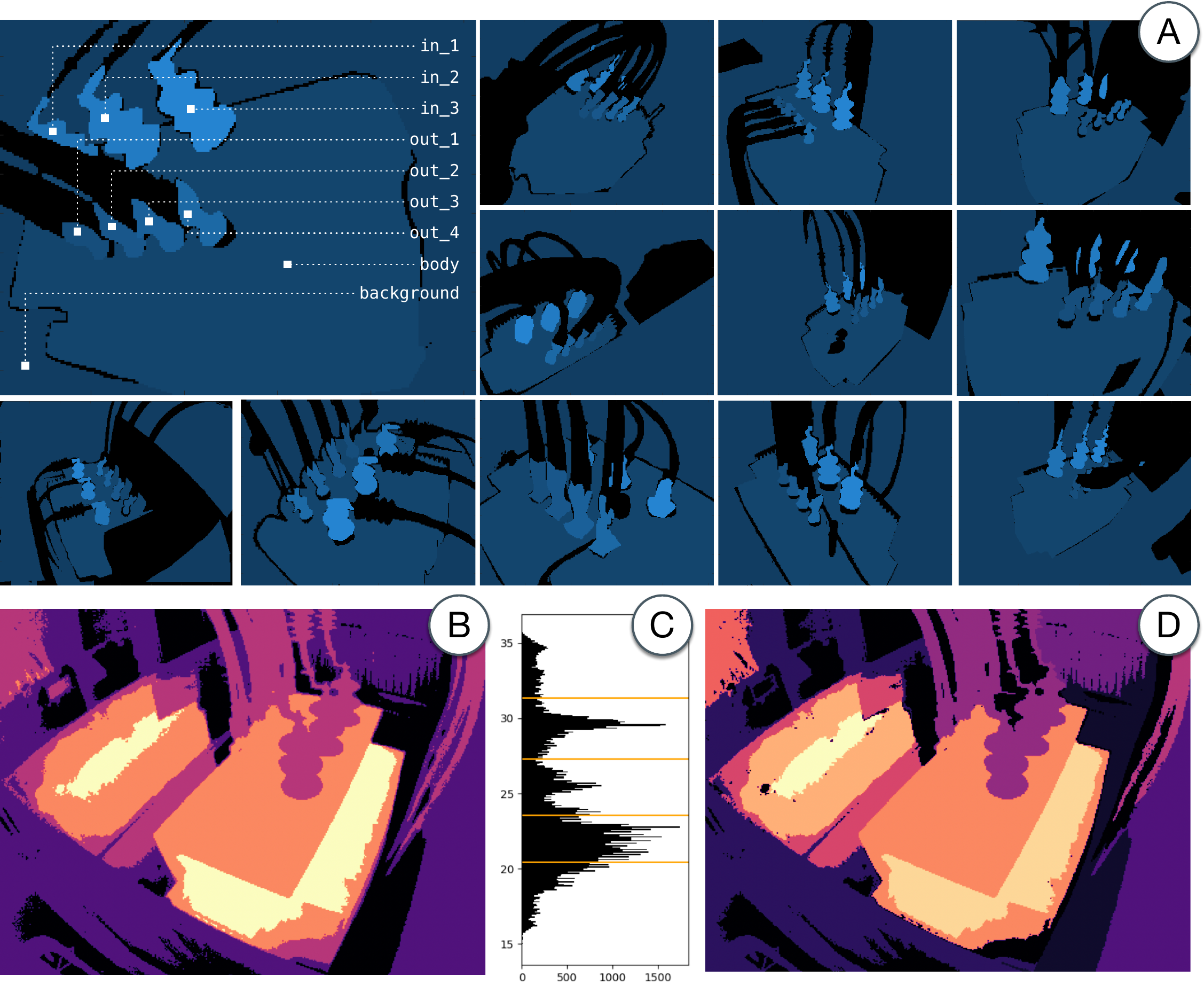}
\caption{\textbf{Manually-defined masks and automatic segmentation results}. (A) Twelve examples of manually-defined segmentation masks are displayed. (B) Five distinct regions automatically detected by Otsu's algorithm. (C) The temperature histogram is depicted alongside the four thresholds used to delineate these five regions. Finally,  (D) depicts the application of the MSER algorithm to each previously identified region, facilitating the identification of spatially separated subregions of equal intensity and resulting in a total of 12 final regions.} \label{masks}
\end{figure}

\paragraph{Manually-defined segmentation masks.} Nine regions of interest (ROI) have been defined from which to extract and monitor their operating temperature. These regions are listed in Table \ref{rois} and include the three primary windings terminals and their high voltage bushings, the four secondary windings terminals and their low voltage bushings, the body of the transformer, and the background temperature (e.g., building walls, excluding cables or other elements that may be a source of heat and could generate false alarms). The thermal cameras have been strategically installed and oriented to capture, as much as possible, the operating temperature of these regions of interest. To extract the temperature of each region, individual masks have been manually-defined for the scene captured by each camera (Figure \ref{masks}). In this way, the operating temperature of a specific region is extracted by computing the arithmetic mean of the top 5\% of pixels with the highest temperature within that region. This process is repeated for the remaining regions in each scene, extracting and appending the operation temperature to the temperature time series. 

\paragraph{Automatic segmentation.} The primary aim of the automatic segmentation process is not to extract the operation temperature of each region, but rather to compute and longitudinally store the sizes of various areas generated by the segmentation algorithms. This data will be utilized to identify anomalies in the scene that extend beyond purely thermal characteristics, such as excessive vegetation growth, the presence of intruders, or fires. Given the fixed positions of the cameras, they capture similar scenes over time. Consequently, the number of pixels belonging to each region automatically generated by the segmentation algorithms should remain relatively consistent across images. Any deviation from this consistency may suggest unusual occurrences warranting further analysis. The automatic segmentation process primarily relies on two segmentation algorithms: the Otsu segmentation algorithm \cite{Otsu1996} and the Maximally Stable Extremal Regions Algorithm (MSER) \cite{Matas2004}. 

The Otsu's method is a non-parametric algorithm widely employed for threshold selection in image segmentation tasks and is characterized by its effectiveness in maximizing the separability of resulting classes in gray levels. This approach hinges on utilizing the zeroth- and first-order cumulative moments of the gray-level histogram, reducing the threshold selection task to an optimization problem aimed at maximizing class separability. Formally, in a scenario with two classes, the threshold $k^*$ is determined as showed in equation 1, where $\sigma _{B}^2$ represents the class separability and can be computed as $ \omega_0 \omega_1 (\mu_1 - \mu_0)^2$, being $\omega_i$ and $\mu_i$ the probability and mean of class $i$.

\begin{equation}
\sigma _{B}^2(k^*) = \max _ {1<k<L }{\sigma _{B}^2(k)}
\end{equation}

Otsu's method offers a means to categorize pixels within an image based on their intensity levels, yet it does not consider the spatial relationships between pixels. In contrast, the Maximally Stable Extremal Regions Algorithm  incorporates this spatial information into the segmentation process. Consequently, regions within the image characterized by similar intensity levels but disjointed spatially (i.e., separated by pixels with differing intensity levels) are identified as distinct regions. This algorithm was originally devised to tackle stereo problems in image analysis (specifically, establishing correspondence between multiple images of the same scene captured from different viewpoints). MSER has found application in various other image analysis tasks, such as tracking colored objects \cite{Donoser2006} or detecting color-based regions \cite{Chavez2011}. A recent work employed this exact methodology to automatically segment parts of an electric power transformer \cite{Segovia2023}. The fundamental concept underlying MSER can be easily described as follows: Imagine a grayscale image represented as a 3D structure, with each point corresponding to a pixel with a height proportional to its intensity level. When a fluid is released onto a point, it spreads outward, encompassing a progressively wider area and incorporating points with lower heights (intensity levels). If this process unfolds simultaneously across multiple points, distinct regions emerge, gradually interconnecting as the fluid level rises. The process terminates when a stability criterion is met, which is continuously checked as the fluid falls.

The proposed solution employed a combination of both algorithms. First, the Otsu's method quantized the registered thermographic image, previously configured to extract nine different regions (multiclass version of the Otsu's algorithm, see Figure \ref{masks}B). Then, the MSER algorithm is applied iteratively to each Otsu's region, thus spatially separated subregions of equal intensity are considered independently (Figure \ref{masks}C). Finally, the number of pixels contained in each region is extracted and stored as time series of region sizes that can be analyzed.

\subsubsection{Prediction model and alarms}

\begin{figure}
\includegraphics[width=\textwidth]{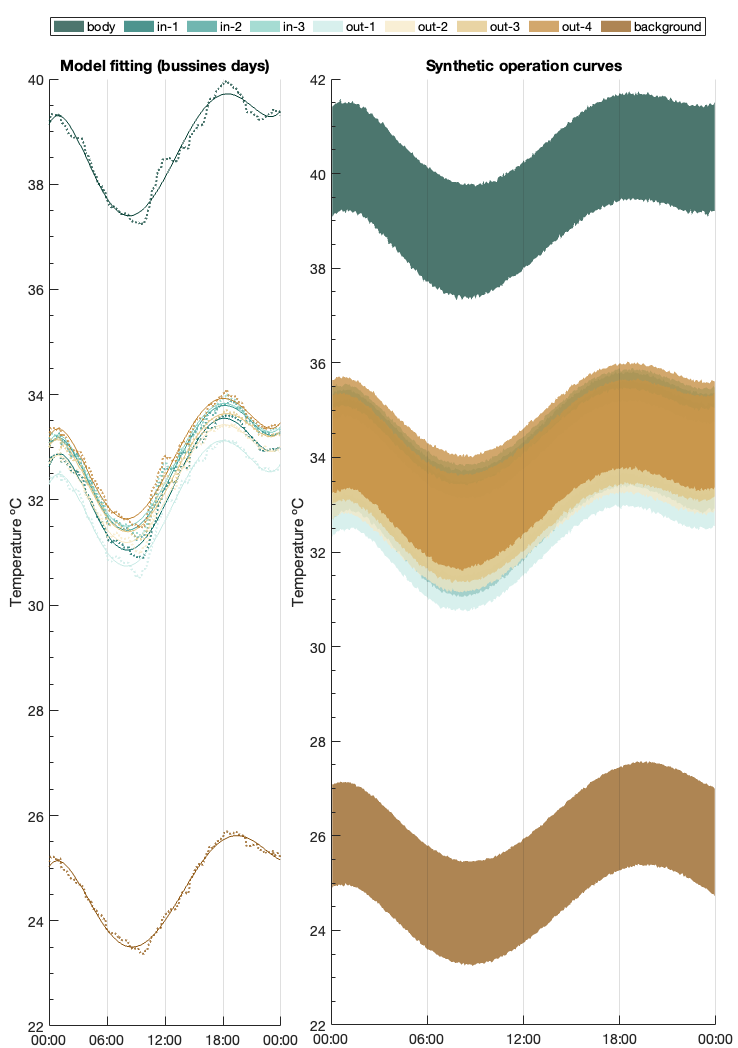}
\caption{\textbf{Synthetic data generation.} (A) This figure displays the fifth-order polynomial curves associated with each region that have been fitted to the actual temperature data model. These curves are subsequently used to extrapolate the transformer's operational curves throughout the entire year. (B) Standard Error of the Mean (SEM) of the 365 synthetic daily curves generated. (C and D) Thermal anomalies generated according to the mathematical model of the thermal system designed for the transformer.} \label{duck_curves}
\end{figure}

Once an image is captured and processed by the segmentation and temperature extraction submodules, the predictions and alarms submodule determines whether the extracted temperatures for each region fall within the expected margins. If not, an anomalous temperature alarm is generated for the specific region and sent to the central server for the operator to evaluate the situation and take appropriate action. These margins are not fixed and depend on various factors such as the time of day, ambient temperature, season, current weather conditions, and the transformer's current workload. It is indeed true that the temperature recorded by the cameras exhibits stationary behavior over time, resulting in daily "duck" curves \ref{duck_curves}. These curves, as will be seen in detail later, depict the transformer's load profile over 24 hours. Due to this observed stationarity, future predictions are calculated adaptively based on past observations, prior to the temperature signal recording. The functioning of the proposed predictive model and the configuration parameters used are explained in detail below.

\paragraph{Auto-Regressive prediction model.} Autoregressive models are a class of simple machine learning models that automatically predict the next element in a sequence of data based on previous elements in that sequence. Autoregression is thus a statistical technique used in time series forecasting that assumes the current value of a time series is a function of its previous values. The number of previous values considered to generate the prediction at the current time is known as the model order (or lag) and is typically denoted by the letter $p$. Thus, an autoregressive model of order $p$ is denoted as $AR(p)$, and the value of the current sample is calculated based on the previous $p$ samples, as indicated in the following equation:

\begin{equation}
	y_t = c + \sum^{i = 1}_{p}{\phi_i y_{t-i}} + \epsilon_t = c + \phi_1 y_{t-1} + \phi_2 y_{t-2} + ... + \phi_p y_{t-p} + \epsilon_t 
\end{equation}

where $c$ is a constant, $\epsilon_t$ is white noise, and $\phi_i$ are the coefficients of the model to be estimated during training. These coefficients represent the importance of each predictor in the final predicted value. The choice of the model order $p$ depends on the amount of information we want to capture to predict future samples. Thus, a higher-order model $p$ will capture more information from the time series than a lower-order one. This might lead us to believe that predictions could be more accurate when we increase the value of $p$. However, this is not always the case; moreover, more complex autoregressive models require higher computational costs. The choice of the model order depends on the nature of the time series data being worked with and must represent a trade-off between capturing enough information to make accurate predictions and the associated computational cost. The proposed model order is individually adjusted for each camera, ensuring that it is always trained with information from the previous 24 hours leading up to the current time point.

\paragraph{Training procedure.} The proposed prediction model is adaptive, meaning its parameters are recurrently updated based on newly available temperature data. This is known as online learning. Such algorithms are suitable for environments where the training data changes rapidly. The choice of retraining frequency is crucial, as the predictive model needs to adapt to the transformer's new daily temperature conditions to make accurate predictions. However, overly frequent retraining should be avoided, as it would incur high computational costs and, more importantly, risk overfitting the model. If a thermal anomalies is occurring and the model is being retrained very frequently (e.g., every time a new temperature value is recorded), it could learn from this anomalous behavior and make predictions accordingly, resulting in a false negative (predicted temperature and recorded anomalous temperature being similar, thus not triggering an alarm).

In the proposed solution, the adaptive prediction model is retrained every 720 minutes (12 hours). As mentioned earlier, the model order will depend on the frequency of thermal image acquisition but will always be configured to use the latest 24 hours for retraining. Thus, the model is trained with the latest 24 hours of available data to predict the temperature for the next 12-hour time window. These predictions, along with the recorded temperature and the size of regions detected by automatic segmentation algorithms, are permanently stored on the server as a time series of data. After the next 12 hours, the system will not have temperature predictions, so they need to be regenerated. To do this, we retrain the predictive model, incorporating the new temperature values recorded in the last 12 hours into the training data, and predict the next time window. This way, as mentioned earlier, the model adapts to the transformer's new temperature conditions and makes predictions accordingly.

\paragraph{Anomaly detection system.} The proposed alarm generation system is straightforward. As long as temperature predictions are available for the time of capturing the current thermal image, the prediction and alarm generation submodule simply needs to check that the difference between the recorded temperature and the estimate does not exceed a predefined threshold (a margin of 15\degree) for each region of interest. Adjusting this threshold allows for tuning the sensitivity of the anomaly detection system. Thus, after acquiring a thermal image, a temporary table is generated containing the recorded temperature data, the expected values for each ROI, and a control bit associated with each ROI indicating the presence or absence of a thermal anomaly alarm. This table information is extracted and sent to the central server through an available API. Upon receipt, the operator can take appropriate action, such as requesting a live video stream to visually inspect the monitored equipment's status. Due to storage space constraints on the server, these temporary tables are periodically deleted as new tables are generated, as they contain redundant information.
\begin{table}[]
\caption{Information table containing the registered temperature, the predicted temperature and a the generated alarm for each region of interest.}
\begin{tabular*}{\textwidth}{@{\extracolsep{\fill}}llllllllll@{}}
\toprule
\multicolumn{1}{l}{} 
&\multicolumn{1}{l}{\texttt{in-1}} 
&\multicolumn{1}{l}{\texttt{in-2}} 
&\multicolumn{1}{l}{\texttt{in-3}} 
&\multicolumn{1}{l}{\texttt{out-1}} 
&\multicolumn{1}{l}{\texttt{out-2}} 
&\multicolumn{1}{l}{\texttt{out-3}} 
&\multicolumn{1}{l}{\texttt{out-4}}
& \multicolumn{1}{l}{\texttt{body}} 
& \multicolumn{1}{l}{\texttt{back}} \\ \midrule
\texttt{temperature} & \texttt{25.78}&\texttt{25.77}&\texttt{25.47}&\texttt{28.03}&\texttt{28.57}&\texttt{28.77}&\texttt{27.21}&\texttt{30.97}&\texttt{21.28} \\
\texttt{prediction} & \texttt{27.39}&\texttt{27.33}&\texttt{27.02}&\texttt{29.66}&\texttt{30.2}&\texttt{30.41}&\texttt{28.83}&\texttt{32.76}&\texttt{23.04} \\
\texttt{alarm} & \texttt{0} & \texttt{0} & \texttt{0} & \texttt{0} & \texttt{0} & \texttt{0} & \texttt{0} & \texttt{0} & \texttt{0} \\
	\bottomrule
\end{tabular*}
\label{api_tables}
\end{table}

\subsection{Synthetic data generation} Since the amount of recorded temperature data is limited, we have decided to generate a sufficiently extensive synthetic database to test the system's long-term performance. This database comprises the daily temperature curves of each of the nine regions of interest selected for analysis (see Table \ref{rois}) over a full year. To achieve this, we have mathematically modeled a thermal system to closely resemble the thermal performance of the transformer.

\subsubsection{Temperature models.} Firstly, we must consider that the energy demand from both industry and network users differs between weekdays and weekends, resulting in distinct temperature curves observed in the data records. Therefore, weekdays and holidays need to be modeled independently. To accomplish this, we selected the temperature profiles $T_{reg}$ from two complete days (2023-10-18 and 2023-10-29) of the real records, which we will use as models to fit the synthetic data we will generate later. The next step is to fit a polynomial curve $T_{pol}$ (of order 5 in this case) to the previously selected real temperature model, repeating this step for each region of interest. In Figure \ref{duck_curves}A, we display the 9 polynomial curves of each region of interest fitted to the real recorded data. Additionally, we assume that the approximation error $e=T_{reg}-T_{pol}$ follows a Gaussian distribution with mean $\mu_e$ and variance $\sigma_e$.  

\begin{table}[]
\caption{Daily temperatures extracted from AEMET.}
\begin{tabular*}{\textwidth}{@{\extracolsep{\fill}}llllllllll@{}}
\toprule
\multicolumn{1}{l}{\texttt{date}} 
&\multicolumn{1}{l}{\texttt{$T_{max}$}} 
&\multicolumn{1}{l}{\texttt{$T_{min}$}} 
&\multicolumn{1}{l}{\texttt{date}} 
&\multicolumn{1}{l}{\texttt{$T_{max}$}} 
&\multicolumn{1}{l}{\texttt{$T_{min}$}} 
&\multicolumn{1}{l}{\texttt{date}} 
&\multicolumn{1}{l}{\texttt{$T_{max}$}}
& \multicolumn{1}{l}{\texttt{$T_{min}$}}  \\ \midrule
\texttt{1-Jan-2022} & \texttt{20.9}&\texttt{0.5}&\texttt{3-May-2022}&\texttt{19.0}&\texttt{9.4}&\texttt{2-Sep-2022}&\texttt{33.1}&\texttt{18.4} \\
\texttt{2-Jan-2022} & \texttt{20.6}&\texttt{-0.4}&\texttt{4-May-2022}&\texttt{17.2}&\texttt{8.5}&\texttt{3-Sep-2022}&\texttt{32.5}&\texttt{16.9} \\
\texttt{3-Jan-2022} & \texttt{20.4} & \texttt{-1.3} & \texttt{5-May-2022} & \texttt{24.0} & \texttt{7.8} & \texttt{4-Sep-2022} & \texttt{33.2} & \texttt{11.8} \\
\texttt{...} & \texttt{...} & \texttt{...} & \texttt{...} & \texttt{...} & \texttt{...} & \texttt{...} & \texttt{...} & \texttt{...} \\
	\bottomrule
\end{tabular*}
\label{aemet}
\end{table}

\subsubsection{Extrapolation for one full year} The temperature recorded by the camera reflects the amount of radiated heat captured by the camera sensor. This captured radiation originates from the emitted radiation of each transformer region as well as the ambient temperature at that time. Thus, in accordance with Stefan-Boltzmann's law, the radiation from the transformer captured by the camera sensor for a specific day could be modeled as:
\begin{equation}
	\gamma_{day} = \alpha \epsilon \sigma[(T_{reg}+273.15)^4-(T_{day}+273.15)^4]
\end{equation}

where $\alpha$ is the camera's reception coefficient, $\epsilon$ is the emissivity of the transformer material, $\sigma$ is the Stefan-Boltzmann constant ($5.67 \cdot 10^{-8} W/m^2K^4$), $T_{reg}$  is the average operating temperature of the model calculated as $(\max({T_{reg}}) + \min({T_{reg}}))/2$ y $T_{day}$ is the average ambient temperature of the day to be modeled, also calculated as $(\max({T_{day}}) + \min({T_{day}}))/2$.\\

Therefore, the synthetic temperature curve for a specific day is modeled as indicated in the following equation, where we introduce the error $e_{day}$, calculated as a sequence of random numbers generated from a Gaussian distribution with mean $\mu_{day}$ and variance $\sigma_{day}$, adjusted to the temperature of the day to be modeled:

\begin{equation}
	T_{op} = e_{day} +[(T_{pol}+273.15)+\gamma_{day}]-273.15
\end{equation}

To model the entire year, we have utilized historical temperature records from the area, stored by the Spanish State Meteorological Agency (AEMET)\footnote{Link to the source: https://opendata.aemet.es/}. These records include daily maximum and minimum temperatures for the entire year 2022, as depicted in the Table \ref{aemet}.

\subsubsection{Generación sintética de anomalías térmicas.} Finally, to the previously generated synthetic temperature records, we added thermal anomalies caused by overheating on random days throughout the year. To simulate such thermal anomalies, we designed a heating function that mimics the heating and cooling of a random region of the transformer over a short period of time. Thus, the temperature increments or decrements over time follow the following nonlinear balance equation:

\begin{equation}
	mc_p\frac{dT}{dt} = UA(T_a-T)+\epsilon\sigma A (T_a^4 - T^4) + \beta Q
\end{equation}

where $m$ es la masa en $kg$, $c_p$ is the specific heat in $J/kgK$, $U$ is the heat transfer coefficient in $W/m^2K$, $A$ is the surface area in $m^2$, $\epsilon$ is the emissivity, $\sigma$ is the Stefan-Bolztman constant in $W/(m^2K^4)$, $\beta$ is the heater factor, $Q$ is the heater output, and y $T_a$ y $T$ are the current temperature and heater temperature respectively in degrees Kelvin. The thermal anomaly generated after apply this heating function over short period of time is depicted in Figure \ref{duck_curves}D.\\

Once we have the temperature time series for the entire year 2022, we need to generate the sequence of corresponding thermal images. That is, for each time point in the series corresponding to the analyzed transformer regions, we generate an image where the pixels belonging to each region have the temperature value at that time point.

\section{Results and discussion} 

\begin{figure}
\includegraphics[width=\textwidth]{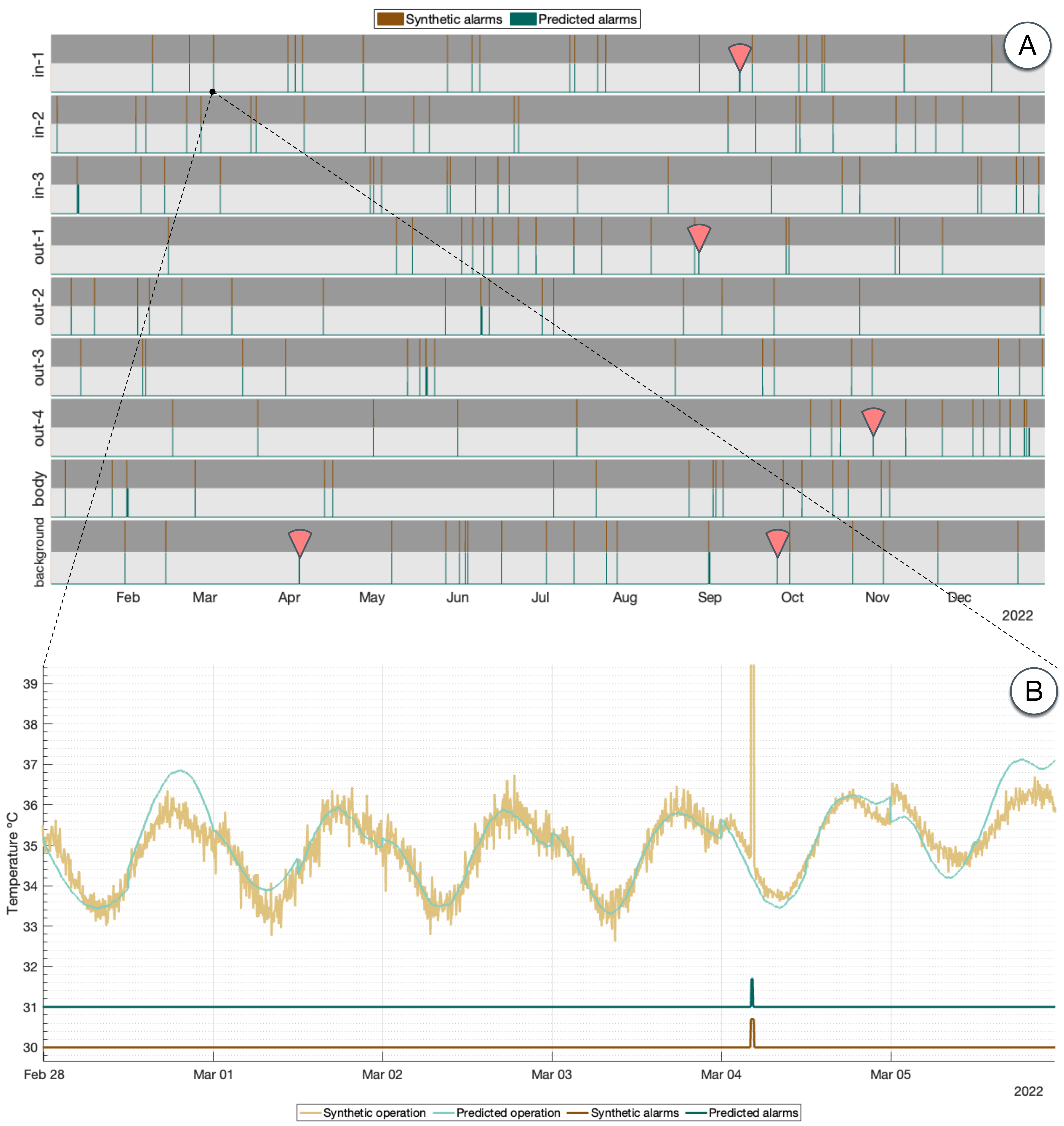}
\caption{\textbf{Simulation results.} (A) This figure juxtaposes the synthetically generated alarms and those predicted by the proposed model after running the simulation script for the synthetic database of the year 2022. False positives are marked with red triangles. (B) Expansion of the previous graph, showing the synthetic temperature values and those predicted by the proposed model. As we can observe, the system correctly detects the thermal anomaly depicted in the image on May 4th in the $in_1$ region.} \label{alarms}
\end{figure}

\subsection{Simulation results} After generating the synthetic database for the entire year 2022, a script was launched to simulate the system's operation throughout that period. To accomplish this, the main program was iteratively and continuously fed with the previously generated synthetic images. This approach enabled the main program to interpret these synthetic images as those collected by the cameras. Accordingly, based on the configuration parameters, it analyzed them, generated historical temperature records, trained the predictive model adaptively, made future temperature predictions, and generated alarms upon detecting a thermal anomaly. The results of this simulation are depicted in Figure \ref{alarms}A, which juxtaposes synthetically introduced thermal anomalies and alarms generated by the prediction model. These results are highly promising, indicating correct detection of the vast majority of anomalies by the system. In specific transformer regions, false positives were observed rarely throughout the year ($in_1$ in September, $out_1$ in August, $out_4$ in October, and $background$ in April and September). These false positives (marked with a red triangle in the figure) can be corrected by adjusting the alarm threshold without compromising anomaly prediction capability. Figure \ref{alarms}B provides a more detailed view of both the temperature time series and predictions made by the proposed model. In this example, the introduced thermal anomaly is accurately detected by the model. Table \ref{acc} displays the temperature deviation calculated as the absolute difference between the temperature predicted by the model and the temperature recorded by the camera throughout the year for each analyzed region. It can be observed that, in the worst-case scenario, the deviation does not exceed 0.6$\pm$4.0 \degree C.

\begin{figure}
\includegraphics[width=\textwidth]{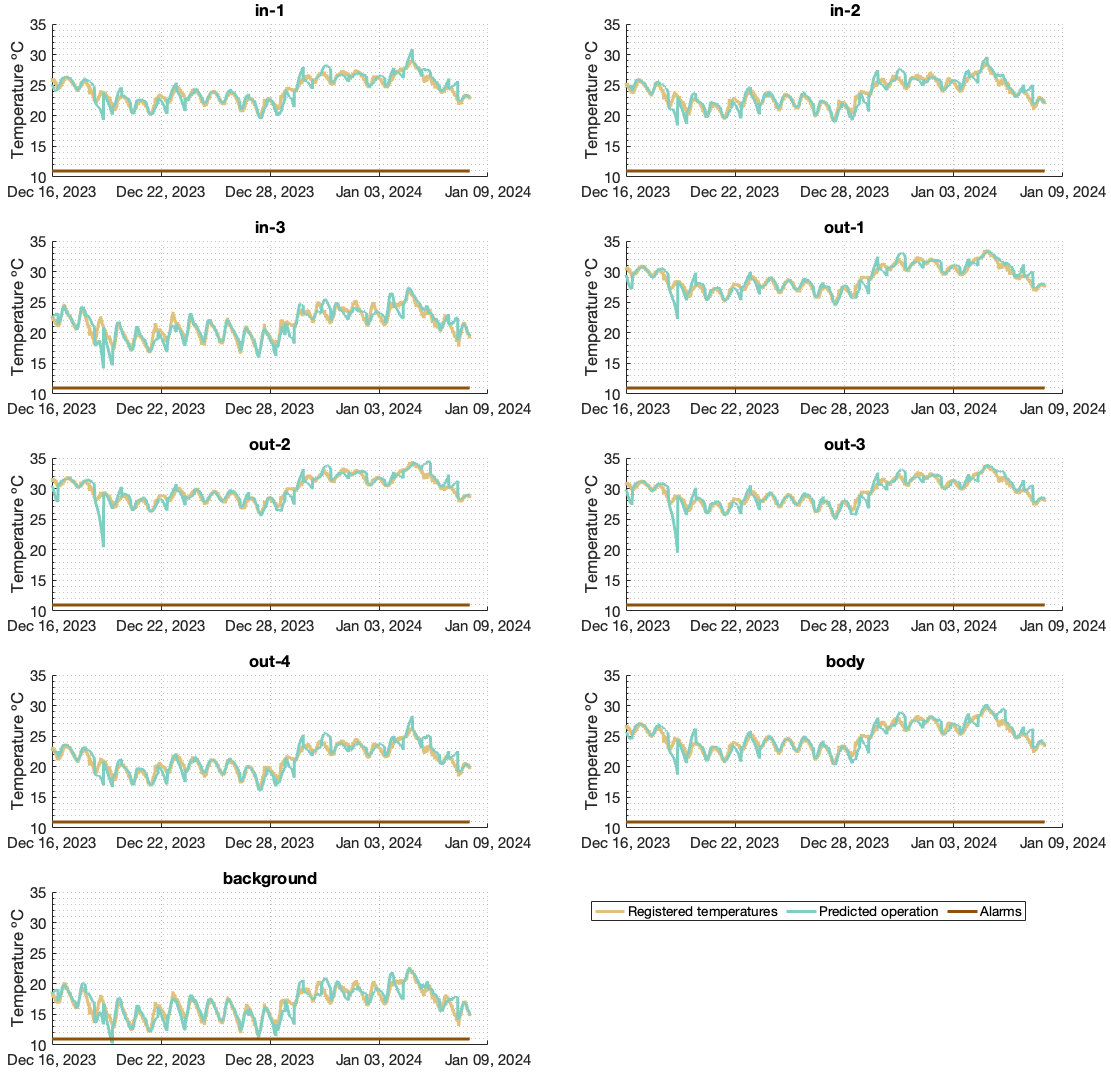}
\caption{\textbf{Registered temperature records}. In this figure, we present the real temperature records obtained from December 16, 2023, to January 7, 2024, in transformer CT49024-master-CATR2. Alongside, the predictions made by the proposed model for all analyzed regions of the transformer are depicted, as well as the record of alarms generated during this period.} \label{data}
\end{figure}

\subsection{Registered temperature time series}
In Figure \ref{data}, we present the actual temperature series recorded from December 16, 2023, to January 7, 2024, for the 9 regions of interest analyzed, alongside the predictions made by the proposed model. As indicated in Table \ref{acc}, the deviation between the actual and predicted temperature values does not exceed 0.81$\pm$1.1 \degree C in the worst-case scenario. Throughout the displayed period, as depicted in the figure, no alarms were triggered in any region. The results obtained from analyzing the remaining thermal cameras are similar to those depicted in this figure.

\begin{table}[]
\caption{Temperature deviations between both real and synthetic records and the predictions made by the proposed model in both cases.}
\begin{tabular*}{\textwidth}{@{\extracolsep{\fill}}llllllllll@{}}
\toprule
\multicolumn{1}{l}{} 
&\multicolumn{1}{l}{\texttt{in-1}} 
&\multicolumn{1}{l}{\texttt{in-2}} 
&\multicolumn{1}{l}{\texttt{in-3}} 
&\multicolumn{1}{l}{\texttt{out-1}} 
&\multicolumn{1}{l}{\texttt{out-2}} 
&\multicolumn{1}{l}{\texttt{out-3}} 
&\multicolumn{1}{l}{\texttt{out-4}}
& \multicolumn{1}{l}{\texttt{body}} 
& \multicolumn{1}{l}{\texttt{back}} \\ \midrule
$\Delta T_{syn}$ & .62$\pm$1.6 & .49$\pm$1.3 & .66$\pm$2.7 & .55$\pm$1.3 & .59$\pm$4 & .56$\pm$2.8 & .63$\pm$1.5 & .67$\pm$2.5 & .59$\pm$4  \\
$\Delta T_{reg}$ & .63$\pm$.87 & .64$\pm$.89 & .76$\pm$1.1 & .58$\pm$.87 & .61$\pm$.99 & .6$\pm$.99 & .66$\pm$.93 & .65$\pm$.94 & .81$\pm$1.1  \\
	\bottomrule
\end{tabular*}
\label{acc}
\end{table}

\section{Conclusions}
In this paper, we introduce one of the core modules of RESISTO, a pioneering technological innovation and research transfer project in Europe. Its primary objective is to enhance the resilience of the electrical grid against meteorological phenomena, thereby improving the quality of service provided to customers in the region. Specifically, we focus on describing the development of a key pillar of the project: the detection of anomalies in the operating temperature of power transformers. To achieve this, a distributed system of servers and strategically positioned thermal cameras has been deployed in the Doñana Natural Park, enabling real-time monitoring of the operating temperature of different regions of interest in the power transformers. The anomaly detection process is divided into three distinct submodules. The segmentation submodule is responsible for automatically segmenting the thermal images captured by the cameras, as well as applying manually defined masks. The temperature extraction submodule extracts and stores the operating temperature of each region of interest. Finally, the anomaly detection and alarm generation system analyzes the recorded operating temperature, predicts the expected operating temperature, and generates alarms if there is a significant deviation between the two. For this latter module, we have developed an adaptive and autoregressive machine learning model capable of predicting the expected operating temperature in real-time for the upcoming hours. To assess the long-term performance of the model, and given the limited availability of real data at present (due to the project's youth), we have generated a synthetic database by mathematically modeling a thermal system similar to that of the transformer. Simulations over a year of synthetic data demonstrate the system's functioning in detecting anomalies in response to sudden changes in the operating temperature in different regions of the transformers. Furthermore, we demonstrate how this system not only operates effectively for synthetic data but also for real data collected by the distributed network of 20 thermal cameras.

\begin{credits}
\subsubsection{\ackname} 
This work is part of the PID2022-137451OB-I00 and PID2022-137629OA-I00 projects, funded by the CIN/AEI/10.13039/501100011033 and by FSE+. Additionally, this work was also supported by the University of Granada and Endesa Distribución under the RESISTO (ref. 2021/C005/00144188) contracts.

\subsubsection{\discintname}
\end{credits}
%
%
%
\bibliographystyle{unsrt}
\bibliography{samplepaper.bib}

\end{document}